\documentclass{article}
\usepackage[latin9]{inputenc}
\usepackage{geometry}
\usepackage{amsmath}
\usepackage{amssymb}
\usepackage[unicode=true,pdfusetitle,
 bookmarks=true,bookmarksnumbered=false,bookmarksopen=false,
 breaklinks=false,pdfborder={0 0 0},backref=section,colorlinks=false]
 {hyperref}

\makeatletter

\DeclareRobustCommand{\greektext}{%
  \fontencoding{LGR}\selectfont\def\encodingdefault{LGR}}
\DeclareRobustCommand{\textgreek}[1]{\leavevmode{\greektext #1}}
\DeclareFontEncoding{LGR}{}{}
\DeclareTextSymbol{\~}{LGR}{126}
\providecommand{\tabularnewline}{\\}

\usepackage{float}
\usepackage{graphicx}
\usepackage{grffile}
\usepackage{breakurl}
\usepackage{caption}

\providecommand{\tabularnewline}{\\}

\usepackage{fullpage}\usepackage[justification=centering]{caption}\usepackage[all]{hypcap}\usepackage{microtype}

\makeatother

\begin{document}

\title{ON SINGLE PARTICLE ENERGIES and NUCLEAR g FACTORS}

\author{ Shadow Robinson$^{1}$and Larry Zamick$^{2}$\\
 \textit{1.Department of Physics}, Millsaps College, Jackson, Mississippi,
39210\\
 \textit{2.Department of Physics and Astronomy}, Rutgers University,
Piscataway, New Jersey 08854 }

\date{\today}

\maketitle
 
\begin{abstract}
If we add one neutron to doubly magic $^{100}$Sn, we can associate
the low lying states in$^{101}$Sn with single particle states. Thus
the J=$\frac{5}{2}^{+}$ and J=$\frac{7}{2}^{+}$states are identified
as d$_{5/2}$ and g$_{7/2}$ states respectively. In $^{101}$Sn,
these two low lying states are separated by an energy of 0.172 MeV.
Currently there is a dispute as to the ordering of these states.We
examine how the 2 scenarios-- selecting J=$\frac{5}{2}^{+}$ as the
ground state or J=$\frac{7}{2}^{+}$ as the ground state-- affect
spectra and nuclear g factors of higher mass Sn isotopes in a variety
of shell model situations. 
\end{abstract}

\section{Introduction}

There is disagreement in the literature about the spin assignments
of the lowest 2 states in $^{101}$ Sn, a nucleus which in the simplest
picture consists of a doubly magic $^{100}$Sn plus one valence neutron.
There is agreement that the splitting of the 2 states is 0.172 MeV,
but whether the ground state has angular momentum J=$\frac{5}{2}^{+}$
and the excited state J=$\frac{7}{2}^{+}$\cite{kat,sew} or vice
versa \cite{darby} is in dispute. In a theoretical work of Jinag
et al. \cite{jiang} they take the ground state to have J=$\frac{7}{2}^{+}$
as per ref. \cite{darby} and the excited state J=$\frac{5}{2}^{+}$.
These states are associated with the single particle neutron orbitals
g$_{7/2}$ and d$_{5/2}$. (The effective shell model interaction
we will use, sn100pn {[}5{]} places the g$_{7/2}$ below the d$_{5/2}$
if one uses the default single particle energies.) In this work we
will keep an open mind about these 2 scenarios and explore the implications
for shell model calculations. We are guided by the observation that
the experimental data for the nearby odd A isotopes of Tin prefer
the picture of a lower d$_{5/2}$. In particular, the odd A Tin nuclei
show that J=$\frac{7}{2}$ is higher than J=$\frac{5}{2}$ for A=103,105,107
and 109. There is a reversal for A=111 and 113 which we will show
is not completely surprising since we are switching from particles
to holes.

In $^{113}$Sn, the ground state is J=$\frac{1}{2}$, J=$\frac{7}{2}^{+}$
is at 77.389 keV while the J=$\frac{5}{2}^{+}$ is at 409.83 keV giving
us a 332.447 keV splitting with J= 5/2 being the higher state.

\begin{table}
\caption{The experimental $J=\frac{7}{2}^{+}$ -- $J=\frac{5}{2}^{+}$ splitting
in the Tin isotopes}
\begin{tabular}{|c|c|cccccc}
\hline 
A  & splitting (keV) & SDI$_{7/2}$  & SDI$_{5/2}$  & SDI$_{0}$  & sn100pn$_{7/2}$  & sn100pn$_{5/2}$  & sn100pn$_{0}$\tabularnewline
 &  &  &  &  &  &  & \tabularnewline
\hline 
\hline 
101  & $\pm$172  & -0.172  & +0.172  & 0  & -0.172  & +0.172  & 0 \tabularnewline
\hline 
103  & 168.0 & -0.008  & +0.055  & +0.033  & 0.244  & 0.283  & 0.279 \tabularnewline
\hline 
105  & 199.7 & -0.024  & 0.046  & 0.005  & 0.258  & 0.347  & 0.305 \tabularnewline
\hline 
107  & 151.2 & +0.015  & -0.001 & 0.014  & 0.264  & 0.246  & 0.246 \tabularnewline
\hline 
109  & 13.38  & 0.003  & -0.037 & -0.025  & 0.212  & 0.103  & 0.177 \tabularnewline
\hline 
111  & -154.48  & 0.031  & -0.094  & -0.011  & -0.038  & -0.284  & -0.164 \tabularnewline
\hline 
113  & -332.447  & -0.008  & -0.352  & -0.018  & -0.656  & -1.0  & -0.829 \tabularnewline
\hline 
\end{tabular}
\end{table}

\section{Calculations. }

We will examine the odd A $^{113}$Sn and the even A isotopes of Sn
in the $g_{7/2}d_{5/2}$ neutron model space.

We begin with $^{113}$Sn. The results are shown in Table 1. We first
examine this nucleus with no interaction, then with the surface delta
interaction (SDI), and lastly with the sn100pn interaction\cite{sn100}.
For each interaction we perform the calculation assuming the $d_{5/2}$
to be the lower shell model orbit (by 0.172 MeV, the splitting in
$^{101}$Sn), g$_{7/2}$ to be the lower orbit, and the case where
the two orbits are degenerate. For each interaction, we will calculate
the $J=\frac{7}{2}$-$J=\frac{5}{2}$ splitting. Experimentally, in
$^{113}$Sn the J=$\frac{7}{2}^{+}$ state is lower than the J= $\frac{5}{2}^{+}$
state by 0.332 MeV while J=$\frac{1}{2}^{+}$ is the ground state.

The no interaction cases serves to show how particle hole inversion
would naturally invert the order of the two low lying states from
their stating point in $^{101}$Sn. The other results show how the
relative spacing is affected when realistic matrix elements are included.
The particle-hole inversion again serves to lower in $^{113}$Sn the
state that was higher in $^{101}$Sn.

How this plays out in nuclei across the low lying Sn isotopes can
be seen in Table 3 where we look at results from the even isotopes
of Sn ranging from A=102 to 110.

As a note, the input splitting in the sn100pn interaction is 0.320
MeV, chosen to get appropriate results in the $^{132}$Sn region.
Changing the splitting to 0.172 MeV as we will do here does not alter
any of the conclusions we will draw. We can use the simple formula
to calculate the change in energies as one goes from $^{101}$Sn to
$^{113}$Sn.

\ensuremath{\Delta}k{o}(j) = -1/(2j+1) {[}\textgreek{S} (2J+1)
\textless{} ( ( j jc) J \textbar{}V\textbar{} (j jc)J - (-1)
(j+jc +J) (jcj )J\textgreater{} {]}. The minus sign is for holes.

The results are shown in Table 2 .

\begin{table}
\caption{Shell Model results for $^{113}$Sn as interaction and lowest orbit
are changed. The splitting between the g$_{7/2}$ and d$_{5/2}$ orbits
is chosen to be 0.172 MeV to match the $J=\frac{7}{2}-J=\frac{5}{2}$
splitting in $^{101}$Sn }
\begin{tabular}{cccccc}
\hline 
 & 7/2 lower  & 5/2 lower  & 7/2 and 5/2 are degenerate in $^{101}$Sn  &  & \tabularnewline
 &  &  &  &  & \tabularnewline
No interaction  & 5/2 is 0.172 lower  & 7/2 is 0.172 lower  & they are degenerate  &  & \tabularnewline
 &  &  &  &  & \tabularnewline
SDI  & 7/2 is 0.008 lower  & 7/2 is 0.352 lower  & 7/2 is 0.18 lower  &  & \tabularnewline
 &  &  &  &  & \tabularnewline
sn100pn  & 7/2 is 0.657 lower  & 7/2 is 1.001 lower  & 7/2 is 0.829 lower  &  & \tabularnewline
\hline 
\end{tabular}
\end{table}

\begin{table}
\caption{The $J=2^{+}$ excitation in the even Tin isotopes}
\begin{tabular}{|c|c|clclclclclcl}
\hline 
A  & Experimental Excitation (keV)  & SDI$_{7/2}$  & SDI$_{5/2}$  & SDI$_{0}$  & sn100pn$_{7/2}$  & sn100pn$_{5/2}$  & sn100pn$_{0}$ &  &  &  &  &  & \tabularnewline
\hline 
\hline 
102  &  & 887 & 876  & 924  & 1230  & 1177  & 1270  &  &  &  &  &  & \tabularnewline
\hline 
104  &  & 815  & 790  & 793  & 890  & 858  & 863  &  &  &  &  &  & \tabularnewline
\hline 
106  &  & 811 & 829  & 820  & 933  & 927  & 933  &  &  &  &  &  & \tabularnewline
\hline 
108  &  & 819  & 813  & 810  & 943  & 1045  & 981  &  &  &  &  &  & \tabularnewline
\hline 
110  &  & 793  & 843  & 817  & 1052  & 1098  & 1090  &  &  &  &  &  & \tabularnewline
\hline 
112  &  & 826  & 884  & 1115  & 1072  & 1091  &  &  &  &  &  &  & \tabularnewline
\hline 
\end{tabular}
\end{table}

In Table3 we show the excitations energy of the lowest 2$^{+}$ of
the Sn isotopes.The near constancy of the energies of J=2$^{+}$ states,
as shown in Table 3 has been noted many times before and was a stimulating
factor in the develpoment of Talmi's Generalized Seniority Model{[}6,7{]}.

\section{Magnetic g factors.}

We can use the work of Yu and Zamick{[}8{]} to calculate magnetic
moments or more precisely nuclear g factors for the even-even Sn isotopes.
Here we however used a matrix diagonalization routine.

We can write the g factor as Ag$_{L}$+ B g$_{S}$. The bare values
of g$_{L}$ and g$_{S}$ for a neutron are respectively 0 and -3.826.We
present results forthe SDI iteraction in Table 4.

We consider bare values, then g$_{L}$= -0.1 g$_{S}$=x {*}(-3.826)
and finally g$_{L}$=+0.1 g$_{S}$=x{*}(-3.826) with the renormalization
chosen to be x=0.75. The negative value of g$_{L}$ is consistent
with meson exchange theory but some have felt the positive value gives
a better fit to the data. The data however is sparse.

In Table 4 we examine the case with no splitting between the $g_{7/2}$
and $d_{5/2}$ orbit. In the no split case it was shown in {[}8{]}that
the g factor is equal to g$_{L}$ for $^{102}$Sn, which in the bare
case is zero.However for $^{104}$Sn and beyond the SDI generates
a splitting of the $g_{7/2}$ and $d_{5/2}$ orbits and so one gets
non-zero g factors. The behavior is somewhat complex. Focusing on
the bare case the g factors which start at zero for $^{102}$Sn become
increasingly negative as one goes to 104 and 106 but the turn around
and become positive by 110, albeit very small (0.0345) . The g factor
only becomes substantially positive for $^{112}$Sn (0.3138).

For the case of the SDI interaction with $g_{7/2}$ orbital being
the lower one, the g factor in $^{112}$Sn is very small because the
splitting J=$\frac{7}{2}^{+}$- J=$\frac{5}{2}^{+}$ is very small
-0.008 MeV as seen in Table 2. If instead we have the J=$\frac{5}{2}$
ground state, the g factor will be larger than the one in Table I,
because the energy splitting is larger (-0.352 MeV). In the no split
case it is only 0.180 MeV.

In tables 5 to 10 we give more detailed results with both the SDI
and Sn100pn interactions.We consider all three scenarios--$d_{5/2}$
ground state in$^{101}$Sn, , $g_{7/2}$ ground state in $^{101}$Sn
and no splititng.we also consider 3 sets of g$_{L}$ and g$_{S}$as
shown. Despite the plethoria of numbers. some general conclusions
can be drawn. Assuming a J =5/2$^{+}$for $^{101}$Sn yields , in
all cases, larger g factors than if we asume a J=7/2$^{+}$ ground
state.. This is undersood ,as per our discussions above about single
particle energies a g factors.In all scenarios the g factor of the
2$^{+}$ state in $^{112}$Sn is larger than the one in $^{110}$
Sn. This seems to go against the experimental trend {[}9,10{]}.The
experimental value of the g factor in $^{112}$Sn is 0.15 with error
bar about 0.05 {[}9{]}. It would appear we can fit this either with
g$_{L}$= -0.1 x=0.75 of the SDI with J=5/2 or g$_{L}$=+0.1 x=0.75
of SDI with J=$\frac{7}{2}$ lower. This serves to underline the interconnected
nature of interaction details and the use of effective g factors.
We will somehow have to have a better understanding of what are the
properly renormalized parameters of the magnetic dipole operator.
As fundamental calculations lead to a negative effective value of
g$_{L}$ for a neutron this parameterization is appealing, but we
cannot say yet that we have a definitive conclusion.

\section{Closing remarks}

This work emphasizes the importance of values of single particle energies
in determining correct g factors in the Tin isotopes. This was also
noted in a schematic calculation by Yu and Zamick{[}8{]} where it
was stated that with degenerate single particle energies and a surface
delta interaction the g factors in $^{112}$Sn vanished. We here focused
on the the J=$\frac{7}{2}^{+}$, J=$\frac{5}{2}^{+}$ energy splittings
in even-odd Tin isotopes. In $^{101}$Sn we identify these as single
particle states 0g$_{7/2}$ and 1d$_{5/2}$, whilst in $^{113}$Sn
these are single hole states. Even in the absence of any interaction
the J=$\frac{7}{2}^{+}$-J=$\frac{5}{2}^{+}$ splitting changes as
we add neutrons to$^{101}$Sn. Indeed in that case the splitting in
$^{113}$Sn is equal and opposite to that in $^{101}$Sn.

We find in all cases the higher the J=$\frac{5}{2}^{+}$ is above
J=$\frac{7}{2}^{+}$ the larger are the g factors of the 2$^{+}$
states in $^{112}$Sn and $^{110}$Sn. There is an ambiguity in $^{101}$Sn
as to whether the J=$\frac{7}{2}^{+}$ state is above or below the
J=$\frac{5}{2}^{+}$ state. If J=$\frac{5}{2}$ (d$_{5/2}$ orbit)
is lower in 101Sn, then due to particle hole inversion, in $^{113}$
Sn the J=$\frac{7}{2}^{+}$state will be farther below J=$\frac{5}{2}^{+}$than
in the scenario where J=$\frac{7}{2}$ ($g_{7/2}$ orbit) is lower.
Thus we will in the former scenario ( g$_{7/2}$ above d$_{5/2}$
) get larger g factors. There is however an ambiguity in this region
of what are the best values of the effective g factors g$_{L}$ and
g$_{S}$ . While logically an effective value g$_{L}$= -0.1 is preferred
the claim has been made that a value g$_{L}$= +0.1 gives a better
fit.

We also considered the 2 measured J=2$^{+}$ g factors g( $^{112}$Sn)
=0.15 {[}9{]} and g($^{110}Sn$)=0.29 {[}10{]}. While one may be able
to adjust one's parameters to fit one of the g-factors, in all the
cases considered we are not able to fit both values. The calculations
give the opposite trend-namely that the g factor of $^{110}$Sn should
be smaller than that of $^{112}$Sn.To clarify the whole situation
it would be of great help if g factor measurements of other isotopes
were were made e.g. $^{108}$ Sn and even lighter Sn isotopes.

\begin{table}
\caption{g factors of the Sn isotopes with the SDI interaction}
\begin{tabular}{|c|c|c|c|c|c|}
\hline 
Sn J=2$^{+}$  & A  & B  & bare  & g$_{L}$= - 0.1 x=0.75  & g$_{L}$=+0.1 x=0.75\tabularnewline
\hline 
\hline 
102  & 1.0000  & 0.0000  & 0.0000  & -0.1000  & +0.1000\tabularnewline
\hline 
104  & 0.9820  & 0.0183  & -0.0699  & -0.1507  & -0.0475\tabularnewline
\hline 
106  & 0.9757  & 0.0244  & -0.0923  & -0.1675  & 0.0276\tabularnewline
\hline 
108  & 0.9809  & 0.0189  & -0.0721  & -0.1523  & 0.0439\tabularnewline
\hline 
110  & 1.0009  & -0.0009  & 0.0345  & -0.0750  & 0.1268\tabularnewline
\hline 
112  & 1.0826  & -0.0826  & 0.3138  & 0.1275  & 0.3439\tabularnewline
\hline 
\end{tabular}
\end{table}

\qquad{}

\begin{table}
\caption{g factors in$^{112}$Sn for various scenarios.}
\begin{tabular}{cccc}
\hline 
$^{112}$Sn  & bare  & g$_{L}$= - 0.1 x=0.75  & g$_{L}$= +0.1 x=0.75\tabularnewline
\hline 
 &  &  & \tabularnewline
splitting 0.172 MeV  &  &  & \tabularnewline
 &  &  & \tabularnewline
SDI$_{5/2}$  & 0.3771  & 0.1729  & 0.3927\tabularnewline
SDI$_{7/2}$  & 0.0110  & -0.0792  & 0.1226\tabularnewline
sn100pn$_{5/2}$  & 0.4216  & 0.2052  & 0.4272 \tabularnewline
sn100pn$_{7/2}$  & 0.4174  & 0.20115  & 0.424 \tabularnewline
\hline 
 &  &  & \tabularnewline
splitting  &  &  & \tabularnewline
 &  &  & \tabularnewline
SDI  & 0.3143  & 0.1275  & 0.34395 \tabularnewline
sn100pn  & 0.4201  & 0.2041  & 0.42605 \tabularnewline
\end{tabular}
\end{table}

\begin{table}
\caption{g factors in$^{110}$Sn for various scenarios.}
\begin{tabular}{cccc}
\hline 
$^{110}$Sn  & bare  & g$_{L}$= - 0.1 x=0.75  & g$_{L}$= +0.1 x=0.75\tabularnewline
\hline 
 &  &  & \tabularnewline
splitting 0.172 MeV  &  &  & \tabularnewline
 &  &  & \tabularnewline
SDI$_{5/2}$  & 0.1748  & 0.026515  & 0.23565 \tabularnewline
SDI$_{7/2}$  & -0.06585  & -0.14765  & 0.04889 \tabularnewline
sn100pn$_{5/2}$  & 0.3516  & 0.1545  & 0.3729 \tabularnewline
sn100pn$_{7/2}$  & 0.1139  & -0.01754  & 0.1884 \tabularnewline
\hline 
 &  &  & \tabularnewline
no splitting  &  &  & \tabularnewline
 &  &  & \tabularnewline
SDI  & 0.03454  & -0.075  & 0.1268 \tabularnewline
sn100pn  & 0.26425  & 0.0913  & 0.3051 \tabularnewline
\end{tabular}
\end{table}

\begin{table}
\caption{g factors in$^{108}$Sn for various scenarios.}
\begin{tabular}{cccc}
\hline 
$^{108}$Sn  & bare  & g$_{L}$= - 0.1 x=0.75  & g$_{L}$= +0.1 x=0.75\tabularnewline
\hline 
 &  &  & \tabularnewline
splitting 0.172 MeV  &  &  & \tabularnewline
 &  &  & \tabularnewline
SDI$_{5/2}$  & --0.03616  & -0.1262  & 0.07195 \tabularnewline
SDI$_{7/2}$  & -0.0923  & -0.1668  & -0.028355 \tabularnewline
sn100pn$_{5/2}$  & -0.1611  & -0.2166  & -0.0250 \tabularnewline
sn100pn$_{7/2}$  & -0.2795  & -0.30235  & -0.11695 \tabularnewline
\hline 
 &  &  & \tabularnewline
no splitting  &  &  & \tabularnewline
 &  &  & \tabularnewline
SDI  & -0.07225  & -0.1523  & 0.0439 \tabularnewline
sn100pn  & -0.2458  & -0.2779  & -0.09075 \tabularnewline
\end{tabular}
\end{table}

\begin{table}
\caption{g factors in$^{106}$Sn for various scenarios.}
\begin{tabular}{cccc}
\hline 
$^{106}$Sn  & bare  & g$_{L}$= - 0.1 x=0.75  & g$_{L}$= +0.1 x=0.75\tabularnewline
\hline 
 &  &  & \tabularnewline
splitting 0.172 MeV  &  &  & \tabularnewline
 &  &  & \tabularnewline
SDI$_{5/2}$  & -0.12265  & -0.18875  & 0.04823 \tabularnewline
SDI$_{7/2}$  & -0.0734  & -0.1531  & 0.043045 \tabularnewline
sn100pn$_{5/2}$  & -0.3673  & -0.36585  & -0.185 \tabularnewline
sn100pn$_{7/2}$  & -0.29985  & -0.31705  & -0.13275 \tabularnewline
\hline 
 &  &  & \tabularnewline
no splitting  &  &  & \tabularnewline
 &  &  & \tabularnewline
SDI  & -0.0933  & -0.16755  & 0.0276 \tabularnewline
sn100pn  & -0.32485  & -0.33515  & -0.17015 \tabularnewline
\end{tabular}
\end{table}

\begin{table}
\caption{g factors in$^{104}$Sn for various scenarios.}
\begin{tabular}{cccc}
\hline 
$^{104}$Sn  & bare  & g$_{L}$= - 0.1 x=0.75  & g$_{L}$= +0.1 x=0.75\tabularnewline
\hline 
 &  &  & \tabularnewline
splitting 0.172 MeV  &  &  & \tabularnewline
 &  &  & \tabularnewline
SDI$_{5/2}$  & -0.1634  & -0.21825  & -0.0268 \tabularnewline
SDI$_{7/2}$  & 0.02912  & -0.0789  & 0.1226 \tabularnewline
sn100pn$_{5/2}$  & -0.33615  & -0.34335  & -0.1609 \tabularnewline
sn100pn$_{7/2}$  & -0.2229  & -0.2614  & -0.0730 \tabularnewline
\hline 
 &  &  & \tabularnewline
no splitting  &  &  & \tabularnewline
 &  &  & \tabularnewline
SDI  & -0.07005  & -0.1507  & 0.04565 \tabularnewline
sn100pn  & -0.28345  & -0.3052  & -0.120 \tabularnewline
\end{tabular}
\end{table}

\begin{table}
\caption{g factors in$^{102}$Sn for various scenarios.}
\begin{tabular}{cccc}
\hline 
$^{102}$Sn  & bare  & g$_{L}$= - 0.1 x=0.75  & g$_{L}$= +0.1 x=0.75\tabularnewline
\hline 
 &  &  & \tabularnewline
splitting 0.172 MeV  &  &  & \tabularnewline
 &  &  & \tabularnewline
SDI$_{5/2}$  & -0.48105  & -0.4482  & -0.27335 \tabularnewline
SDI$_{7/2}$  & 0.308  & 0.1230  & 0.339 \tabularnewline
sn100pn$_{5/2}$  & -0.581  & -0.5205  & -0.3510 \tabularnewline
sn100pn$_{7/2}$  & 0.3585  & 0.1595  & 0.37825 \tabularnewline
\hline 
 &  &  & \tabularnewline
no splitting  &  &  & \tabularnewline
 &  &  & \tabularnewline
SDI  & 0  & -0.1  & 0.1 \tabularnewline
sn100pn  & 0.05785  & -0.05805  & 0.1450 \tabularnewline
\end{tabular}
\end{table}

\end{document}